# THE EFFECT OF SOCIAL WELFARE SYSTEM BASED ON THE COMPLEX NETWORK

Dongwei Guo, Shasha Wang, Zhibo Wei, Siwen Wang and Yan Hong

Department of Computer Science and Technology , Jinli University, Changchun City, China
wangss9219@163.com

## ABSTRACT

*With the passage of time, the development of communication technology and transportation broke the isolation among people. Relationship tends to be complicated, pluralism, dynamism. In the network where interpersonal relationship and evolved complex net based on game theory work serve respectively as foundation architecture and theoretical model, with the combination of game theory and regard public welfare as influencing factor, we artificially initialize that closed network system. Through continual loop operation of the program , we summarize the changing rule of the cooperative behavior in the interpersonal relationship, so that we can analyze the policies about welfare system about whole network and the relationship of frequency of betrayal in cooperative behavior. Most analytical data come from some simple investigations and some estimates based on internet and environment and the study put emphasis on simulating social network and analyze influence of social welfare system on Cooperative Behavio .*

## 1. INTRODUCTION

At present, the game theory is widely applied in many fields such as economics, sociology and computer science .Scientists in various fields have put forward applicable game models, such as snowdrift game, coordination game and so on, and had good simulation results[1]. The prisoner's dilemma model is a typical model in game theory, which interpret the process of game simple and clear. But the prisoner's dilemma model show that individuals choose betrayal will get the highest interest, and betrayal of dominant. However, this does not accord with actual situation, most people need cooperation to complete something in reality. And only under the common cooperation and the long-term relationship, the individual will achieve maximum benefit. At the same time, because the real world not only have the individual game, but also usually a game between people. So people often research multi-player games[2].

Complex network[3] is the research focus in the academic field in the 21st century, and it is used in various fields to describe the relationships between all kinds of complicated things[4-5]. In today's society, the evolution of social welfare system in the developed countries, mainly experienced from residual welfare model to system of welfare model to hybrid welfare model transition[6], and China's social welfare system still exist many problems, and still in the process of perfecting. At present, most studies of social welfare system is in view of economics, there are also many people made comparison and research between structures of national welfare system, but no one study the application of complex network to the social welfare system. Today's society has formed a complex network with the person node. In this complex network, the social welfare system for the cooperative behavior of people will produce what kind of impact? In this paper, we combined with a complex network game model, simulate the behavior of the social individual performance through relevant features to analysis of network status.

This paper will use the complex network to set up a social group with some orders of magnitude. Under the natural development, Between people will influence each other and the implementation of the welfare system and changes will affect most people in this group. when everyone is affected and will feedback to the welfare system and affect other people around us, the people affected with the cooperation of others will also change, may betray partners, give up cooperation, and so on. In this big system of the dynamic and changeable, there are always Effect and feedback. We will study such a system, and establish a model. In this model, the node represents the individual,and edge is abstracted as the relationship between the individual and eventually form a dynamic model, to reflect the change of the whole system. And analysis how the social welfare system based on the complex network effect on the cooperative behaviors. In this paper, we found that let all individuals in the system to balance as soon as possible by dynamically adjusting tax rate and the minimum guarantee of the welfare center value, and let the wave amplitude decreases, and the network of betrayal of inhibition.

## 2. THE INTRODUCTION OF RELEVANT CONCEPTS

Welfare system corresponding to the social group is a social network, social network and other natural network is also a complex system, from the social network also found the famous six degrees of separation phenomenon (also known as the 'small world effect).In this paper we use the characteristics of small world network[2] to established the network model.

Welfare center: responsible for the management of the entire network, establish welfare system (including the tax rate and poor issuance rate) of the department.

Person: the individual in the network interact with each other, with the attribute of personal id, personal wealth, risk factor and the list of the contact.

Risk factor(SPIRIT): An individual's definition to investors, The greater the risk coefficient of individuals tend to invest more,($0 \leq SPIRIT \leq 100$).

The contact list: private property, divided into red list and white list. If the times of successful cooperation higher than that of the number of failed, we put it into the red list, whereas in the white list.

The blacklist: the list of the system, all people share, depositing a betrayal of people.

Rate(η): the proportion of the object charged.If the cooperation with the partner had been success, you must pay income tax according to the specified rate.

Poor issuance rate: the proportion of the funding for poor person.Refers to the property below the lowest life, welfare center according to its property status, issuance rate subsidies granted by poverty.

Earnings ratio: Profit ratio when the successful cooperation.

Percentage: the ratio When failed in the cooperation.

The minimum level of consumption(min): The minimum balance individual normal life.When individual wealth is lower than this value, welfare center will be given the allowance according to the level of poverty.

General level of consumption(aver): the general consumption of the individual's normal life. In this article, we supposed that each traverse round the network through the time t, then every time t, the property of the general individual consumption value is Aver.

Minimum value: the lowest wealth welfare center give to the poor people.

Quick deduction ($\delta$): quick deduction =the next higher level of the highest income×(the rate at this level -The higher tax rate) + the higher quick deduction.

## 3. THE INTRODUCTION OF THE SMALL WORD NETWORK MODEL

This paper uses the small world network model[2] to establish the network. Everyone can be regarded as nodes in complex networks. There are a large number of edges connecting to the network nodes, and two nodes connected by a line are regarded as persons who know each other. We assume that the network had N nodes, set up steps as follows:

Step 1: Each node should be connected to K / 2 neighbor nodes around (K is even), forming the regular net. We number N nodes of network for 1 ~ N and K lines connected to the node for 1~K.

Step 2: Whether Line 1 of node 1 would be reconnected depends on probability p (0<p<1). Node 1 of line 1 stays the same and on the other end is reconnected to other nodes of network. It meets only a line between two points.

Step 3: Line 1 of node 2 to node N repeats step 2 until finished.

Step 4: Line 2 of node 1 to node N repeats step 2 until finished.

Step 5: Processing Line 3 to Line K of node 1 to node N by iterating until finished.

The line will never happen reconnecting while p=0 and it will be homogeneous network. The line must be a reconnect while p=1 and it will be random network. By the introduction of Watts and Strogatz ' small world network[2] in 1998, we know that small world network have characteristic of short average path length and high degree of clustering. Let's call degree of clustering C(p) and average path length $I_{avg}(P)$. Figure 1 shows degree of clustering and average path length is reduced with the increase of p value, but the decreasing degree is different. It can make degree of clustering high and average path length short while p=0.1, so this paper make p=0.1 to construct small world network.

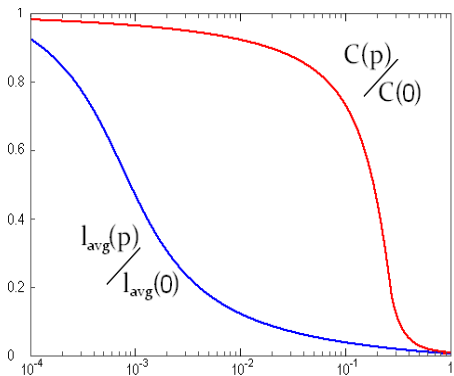

Figure 1. The image of degree of clustering C (red) and average path length L (blue) change with p.

# 4. THE THEORETICAL MODEL

The line will never happen reconnecting while p=0 and it will be homogeneous network. This paper is based on a small-world network to simulate real network. In the network, we define a welfare institution and many individuals. The welfare institution manages the entire network, so it sets rate and poor fund-issuance rate. Through simulate individual income tax in the real society, the welfare institution taxes on profitable individuals from cooperation. If the profit is lower than the threshold, the welfare institution wouldn't tax. Otherwise it taxes according to the following way

tax = taxable income × applicable tax rate quick calculation – deduction

Each time t, the welfare institution statistics its own wealth value and the wealth value it should give out the poor people next time. When the former is lower than the latter, it will reduce the minimum guarantee value. So this will reduce distribution of wealth value of the poor individual and raise tax. Otherwise, if the former exceeds the latter over the continuous time 10t, it will reduce tax and raise the minimum guarantee value. This paper starts with the minimum guarantee value to divide poverty levels into three levels, and set different poor fund-issuance rate with the different poverty levels. Poor fund-issuance rate is unchanged, and distribution of wealth value of the poor individual is linearity to changes of the minimum guarantee value.

In the model, we suppose that the interaction between individuals is through investment a comment project. Investment projects will have the risk, so according to the related factors, it may succeed or fail. When RISK changes, GAIN and LOST along with the change. We suppose M cooperate the same project, so

$$RISK = \sum SPIRIT_{项目参与人} / M + \omega_1$$

$$GAIN = (RISK + \omega_2) \times 0.01$$

$$LOST = (RISK + \omega_3 \% (RISK+1)) \times 0.005$$

Figure 2. Formula 1

Among the above, the range of ω1 which is a random number is [-10,10], and the range of ω2 which is a random number is [0,100], and ω3 is a random number.

In this paper, we will combine the multi-player games model with single-player game. A person can choose to cooperate with other persons who have a connection with him, ask others to invest a project, or he can invest himself. However, individual investment would take bigger risks. In the multi-player game, Sponsor can ask all persons who exist in the red list, part of the white list for investing a project. But the requested people must ensure that their assets is greater than the half of sponsor. Then the requested people can decide whether agree to cooperate by their policy set. In the network, everyone has two alternative strategies,investment and non-investment. In the process of cooperation, μi represents the budget investment of Node i, Pi represents expected profit. Through the value of Pi determine whether agree to cooperation. When Pi > 0, individuals decide to invest, otherwise individuals declined to invest. α Represents the probability of success,

$$Pi = \alpha \times (\mu_i \times GAIN \times (1-\eta) + \delta) - \mu_i \times (1-\alpha) \times LOST$$

Figure 3. Formula 2

In order to maximize their own interests, individuals by their own the value of risk and the investment situation to make their own decisions. Whether individual invest and how invest is called the game process. The success of the investment is determined by probability. We generate a random number between [0,100] compared with value at risk, when its value is greater than the risk, investment success. On the other hand, investment failure. According to investment value and earnings ratio (or loss ratio)of each investor in the project, each investor charges (or pays) the corresponding amount. If investors cooperation successfully, investors will update their contact list. They will put the investors they do not know before in their white list, make investment cooperation times of investors who belong to red list to increase 1, put the investors belonging to white list in red list and delete them from white list. If investors cooperate unsuccessfully, they will make investment cooperation times of investors who belong to red list to decrease 1. What is more, if investment cooperation times of investors who belong to red list are equal to 1, they will put the investors belonging to red list in white list and delete them from red list. However, in the process of investment a project, it may exist individual betrayal, namely , the individual will take all the money investors invest away. Individual will be punished if they choose to betrayal. In a period of time T (T = kt (k > 10)), it exists in the black list, individuals in the network will not cooperate with people in the black list, and people cannot be a single investment in black list. After a period of time, betrayers cannot cooperate with others that lead to betrayers can only consume without income. When there is not enough wealth individuals supported the $\gamma *t$ time consumption, If individual ensure the money of investment is greater than its $\beta$ times the cost of investment and can support in T time consumption and can ensure the budget surplus amount S is enough for their minimum consumption in the next T time, they will choose to betray. So an individual judge whether betrayal by following condition:

① $value(i) < \gamma \times \overline{aver}$

② $Total\_value \geq \beta \times value(i)$

③ $S_i = (Total\_value + Self\_value(i)) - (\theta \times \overline{aver}) \times k$

$S_i > min;$

Figure 4. Formula 3

Total_value means the total wealth value of investment. value(i) means the investment cost of an individual, Self_value(i) means wealth value of an individual. Si means surplus wealth value after an individual is deleted from the black list. θ means positive integer.

Group of betrayal by following condition:

① $Max\{\overline{value(i)}\} < \gamma \times \overline{aver}$ (i=1, 2…d)

② $Total\_value \geq \beta \times \sum_{1 \leq i \leq t} value(i)$

③ $S = (Total\_value + \sum_{1 \leq i \leq t} Self\_value(i)) - (\theta \times \overline{aver} \times d) \times k$

$S > min \times d$

Figure 5. Formula 4

d means d people choose to betrayal.

Each time after t, according to relation factor, welfare center adjusts the tax rate and the lowest guarantee value. Society's total wealth and welfare.

## 5. MODEL ANALYSIS

Network system was closed, in a closed system, Figure 6 show the interaction relationship:

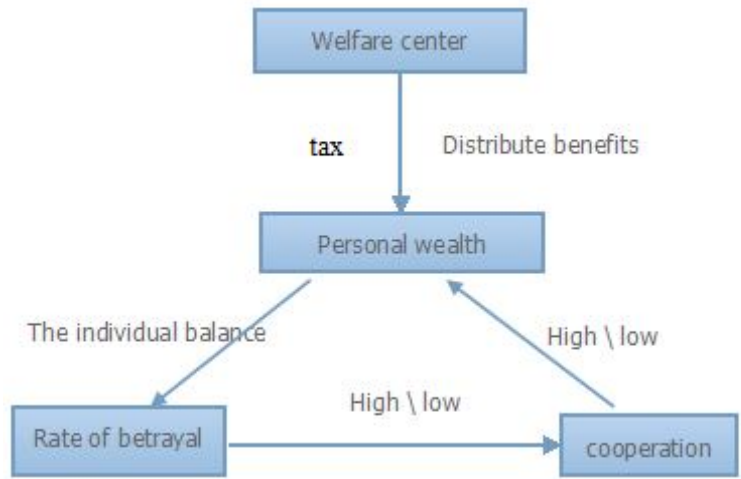

Figure 6  Relational model

(1) Balance of personal payments determines the betrayal, if the estimated income is lower than spending, the treachery must happened in the individual cooperation, and use the betrayal income to make up the difference between income and expenditure. If the estimated income is higher than spending, There is little possibility of individual betrayal.

(2) Betrayal can affect the cooperation of the result, When higher rate of betrayal of the whole system, all the individuals will worry about the partner's betrayal, and they select to reduce the cooperation or don't cooperate with others to avoid risk. On the contrary, if the times of the betrayal is less, less individual in cooperating with other individuals would be worried about the betrayal of partners and the possibility of cooperation will rise.

(3) Presence of ontogenesis of treason in the cooperation, cooperation success or failure will directly affect the individuals involved in cooperation of balance of payments. Failure and partner of betrayal will reduce other individual income, which can let the part of individual income lower than spending.

The above three relations influence each other and restrict each other. When one of them change and make the balance of the whole system destroyed, the other two relations will change accordingly with interdependent function. But the system will balance again, which is the ability of self-regulation. In a word, the whole system vary like like is the cosine function and float up or down on both sides of the line of balance. But depend on the adjustment ability by the system itself, the amplitude of fluctuation is larger. Through the study we found that the system of all individual can achieve the balance of payments as soon as possible by dynamically adjusting the influence factors of welfare center, which is the welfare system. In this way, it can quickly close to the line of balance and make fluctuation amplitude decreases. So reasonable welfare system can make the system to stabilize as soon as possible, keeping society stable state.If the welfare system is not reasonable or too light, it may have not good effect. If the welfare of the welfare center is too much, it may be effective in a short time. But can cause negative effects after some time, and make a lot of individuals depend on the welfare system and not participate in cooperation, which make the welfare system's pressure too large,and its benefits decrease. At this time,the rate of betrayal will rebound again. Figure 7 shows the relationship between them.

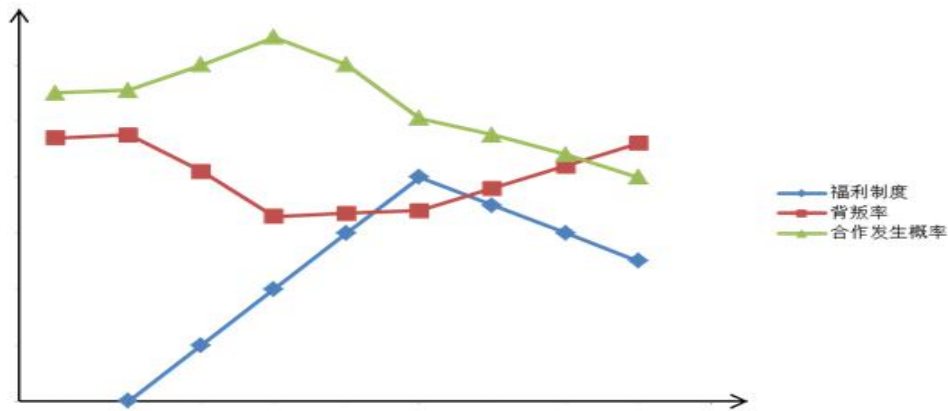

Figure 7   The influence of the relationship between factors

In this picture we can see that there is no influence of the welfare system in the first stage, and the betrayal rate of the system and the possibility of cooperation in near the equilibrium value. The welfare system join in in the second stage. The betray in the cooperation will reduce affected by the welfare system. Because of the reduce of the betray, the possibility of the cooperation may get higher. In the third stage, although the welfare system continues to increase, the rate of the betrayal reach a steady state instead of continuing to decline, the possibility of cooperation began to fall slowly.This is due to the influence of betrayal rate in cooperation reduced or even disappear, but the good welfare system make some individuals give up to participate in the cooperation and depend on the welfare system.In the fourth stage, because of the magnitude of the welfare system to achieve perfection and more people depend on the welfare system, the welfare system was forced to cut, which can cause betrayal rate rebound and the possibility of the cooperation will continue to decline. So a good welfare system can make the system to achieve balance as soon as possible, and make the system stable relatively.

## 6. CONCLUSIONS

Welfare system can affect the speed to the balance and the amplitude of fluctuations of the whole. Reasonable welfare system can make the overall balance quickly, reduce the amplitude of fluctuations, and contributes to the development of the whole. On the contrary,if the welfare system is not reasonable, it will lead to the change among the cooperation in the system larger or smaller impact on the system.

## ACKNOWLEDGEMENTS

This work was supported by the national training programs of innovation for undergraduates in Jilin university. Then we would like to extend our sincere  gratitude to our tutor Dongwei Guo for his helpful guidance and instructive suggestions.

## REFERENCES

[1] Richard Swedberg, (2001) "Sociology and game theory: Contemporary and historical perspectives", Theory and Society, Vol. 30, No. 3, pp301-335.

[2] Liang Chen, Shiqun Zhu, (2008) "This is my paper", Journal of Suzhou University(Natural Science Edition), Vol. 24, No. 3, pp55-59.

[3] Watts DJ, Strogatz SH, (1998) "Collective dynamics of small-world networks", NATURE, Vol. 393, No. 6684, pp440-442.

[4] Lei Guo, Xiaoming Xu, (2006) *Complex Networks*, Shanghai Scientific & Technological Education Publishing House.

[5] Yongkui Liu, (2010) Study of Complex Networks and Evolutionary Game Dynamics on Networks [D], Xi'an: Xidian University.

[6] Yingsheng Li, Yangdi Han, Yifan Xiao, Ning Zhang, (2007) "Beyond the integration model:Classification of urban subsistence allowances system reform problem research", Academia Bimestris, No. 2, pp114-123.

**Authors**

Short Biography: Dongwei Guo was born in 1972.professor, Jilin university Ph.D. His main research interests include knowledge engineering and expert systems.

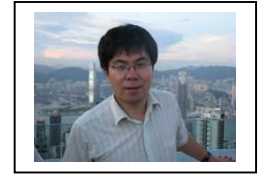

Short Biography: Shasha Wang was born in Hebei province of China, on Sept. 19, 1993. She is a senior student of Jilin University. Now her major is network and information security.

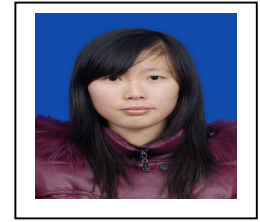

Short Biography: Zhibo Wei was born in Jilin province of China, on Apr. 19, 1993. She is a senior student of Jilin University. Now her major is network and information security.

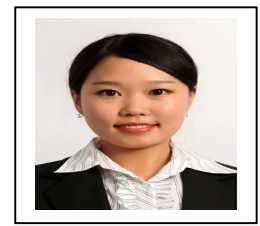

Short Biography: Siwen Wang was born in Jilin province of China, on Apr. 14, 1993. She is a senior student of Jilin University. Now her major is network and information security.

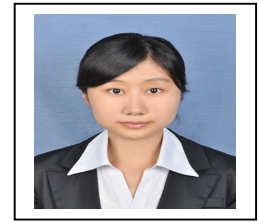

Short Biography: Yan Hong was born in Shaanxi province of China, on Jan. 27, 1992. He is a senior student of Jilin University. Now his major is network and information security.

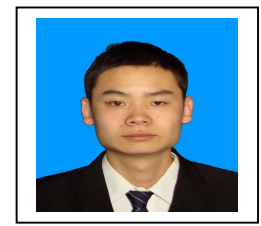